\documentclass[letterpaper, 10 pt, conference]{IEEEconf}

\usepackage{amsmath,amssymb,amsfonts,epsf,epsfig,times}
\usepackage[all]{xy}
\usepackage{graphicx,color}
\usepackage{subfigure}
\usepackage{url}
\usepackage{cite}
\usepackage{tikz}
\usepackage{epstopdf}
\usepackage{algorithm}
\usepackage[noend]{algpseudocode}

\usepackage{keyval,times}

\newtheorem{theorem}{Theorem}[section]
\newtheorem{lemma}{Lemma}[section]
\def\proof{\noindent{\it Proof: }}
\def\QED{\mbox{\rule[0pt]{1.5ex}{1.5ex}}}
\def\endproof{\hspace*{\fill}~\QED\par\endtrivlist\unskip}
\newcommand{\re}{\mathbb{R}}

\newcommand{\norm}[1]{\left\|#1\right\|}

\newcommand{\defeq}{\stackrel{\triangle}{=}}

\newtheorem{corollary}[theorem]{Corollary}

\newcommand{\Lcal}{\mathcal{L}}

\newcommand{\Ncal}{\mathcal{N}}
\newcommand{\Ecal}{\mathcal{E}}
\newcommand{\Gcal}{\mathcal{G}}

\newcommand{\Vcal}{\mathcal{V}}

\newcommand{\OMIT}[1]{}

\makeatletter
\def\BState{\State\hskip-\ALG@thistlm}
\makeatother

\begin{document}
\title{Towards Energy-Efficient Communication Management in the Distributed Control of Networked Cyber-Physical Systems}
\author{Yongcan Cao and Eduardo Pasiliao
\thanks{Yongcan Cao is with the Department of Electrical and Computer Engineering, The University of Texas, San Antonio, TX 78249. Eduardo Pasiliao is with the Munitions Directorate, Air Force Research Laboratory, Eglin AFB, FL 32542.  
}
}

\markboth{}
         {}

\maketitle

\begin{abstract}
In this paper, we study the distributed control of networked cyber-physical systems when a much more energy-efficient distributed communication management strategy is proposed to solve the well-studied consensus problem. In contrast to the existing potential-based network topology control method, the proposed topology control method is based on the variation of communication ranges such that each agent can control its \textit{ad hoc} communication range. The proposed network topology control technique can not only guarantee network connectivity but also reduce the communication energy. We apply the new network topology control technique, based on variable communication ranges, in a well-studied consensus problem, where the communication range for each agent is designed locally along with a new bounded control algorithm. Theoretical analysis is then provided to show that the proposed network topology control technique can guarantee consensus with \textit{bounded} communication energy consumption. Finally, simulation examples are provided to show the effectiveness of the proposed \textit{energy-efficient} distributed topology control technique.    
\end{abstract}


\section{Introduction} \label{sec:Introduction}

The increasing interconnection of physical systems through wireless networks has been observed in different areas, such as sensor networks~\cite{AkyildizSSC02}, unmanned systems~\cite{CasbeerKBM06}, and transportation networks~\cite{NegenbornSchutterHellendoorn08}. One critical issue in the networked cyber-physical systems is the connectivity issue when physical systems need to maintain ``sufficient'' information exchange in order to accomplish the desired team mission. 

To deal with the connectivity issue, one common approach in the control systems design is to introduce artificial potentials that characterize the relative distances between agent pairs~\cite{JiEgerstedt07,ZavlanosPappas07,ZavlanosPappas08,DimarogonasKyriakopoulos08,SuWangChen10,CaoRen12,KanDSD12,PoonawalaSES15}. The artificial potential between a pair of agents is designed in such a way that it will grow to be sufficiently large (could be unbounded) when the distance between them increases to be equal to the communication range. When the control algorithms are designed based on the sum of the gradients of the artificial potentials, the total artificial potential is thus nonincreasing. This then indicates that the initial communication patterns can be preserved because otherwise the total potential will become larger than the initial total artificial potential, as soon as some communication pattern is broken. Such a technique has been used to solve formation control/tracking~\cite{ZavlanosPappas07,ZavlanosPappas08,DimarogonasKyriakopoulos08,CaoRen12,KanDSD12,PoonawalaSES15} and consensus~\cite{JiEgerstedt07,SuWangChen10,CaoRen12}. Although this approach provides a systematic way to guarantee connectivity, the corresponding control algorithms may require arbitrarily large control inputs and a significant amount of communication energy, which is impractical in real-world applications. Other than this disadvantage, the potential-based network topology control technique can only be used for undirected communication in the continuous-time setting.

To address the need for more energy-efficient and practical network topology control in networked cyber-physical systems, we here propose a new approach based on variable communication ranges, when each agent has limited \textit{but variable} communication ranges. The main idea is to change the communication range of each agent as needed to ensure that the desired communication pattern can be preserved. There are two main reasons that we consider variable communication ranges. First, the control input design and network topology control can be decoupled such that the control system design becomes easier. Second, more energy-efficient management of communication resources can be accomplished through adaptively adjusting the communication ranges. 

The contributions of our study are threefold. First, the proposed variable communication ranges can be used for networked systems with direct interaction graphs in the discrete-time setting. The existing potential-based connectivity control techniques can only be used for undirected graphs in the continuous-time setting. Second, bounded and state-independent control input is needed to ensure network topology control. The existing potential-based connectivity control technique requires the adjustment of control input appropriately based on the current states. In many cases, very large control inputs are needed to maintain desired connectivity. Third, much less communication energy is needed than the traditional approach. Since the proposed variable communication ranges take into consider the \textit{value} of communication ranges in real time, adjustment of communication ranges can reduce communication energy consumption without sacrificing team performance. 

The remainder of the paper is organized as follows. Section~\ref{sec:GT} briefly reviews the graph theory notations used throughout the paper. The problem to be studied in this paper is then described in Section~\ref{sec:PF}. Section~\ref{sec:consensus_1st_order} is the main body of the paper that presents the control algorithm design, variable communication range design, and the stability analysis. This section also includes further analysis on the communication energy consumption and its comparison with the traditional approach. Section~\ref{sec:simulation} provides some simulation examples to illustrate the effectiveness of the proposed technique. A short conclusion is given in Section~\ref{sec:conclusion} to summarize the contributions of the paper.


\section{Graph Theory}\label{sec:GT}
For a team of $N$ sensors (also referred as agents for generality), their interaction can be described by a directed graph $\Gcal\defeq (\Vcal,\Ecal)$, where $\Vcal=\{1,\cdots,N\}$ is the agent set and $\Ecal=\Vcal^2$ is the edge set. An edge in a directed graph $\Gcal$ denoted as $(i, j)$ means that agent $j$ can obtain information from agent $i$ (but not necessarily vice versa). That is, agent $i$ is a neighbor of agent $j$. We use $\Ncal_j$ to denote the neighbor set of agent $j$. A directed path is a sequence of edges of the form $(v_1, v_2), (v_2, v_3), . . . ,$ where $v_i\in \Vcal$. A directed graph has a directed spanning tree if there exists at least one agent, also referred to as \textit{a root}, that has directed paths to all other agents.

For a directed graph, we can also use a row stochastic matrix $A=[a_{ij}]\in\re^{N\times N}$ to describe it. A row stochastic matrix is a square matrix whose entries are all nonnegative and the sum of each row is $1$. In particular, $a_{ij}>0$ if $(j,i)\in\Ecal$ and $a_{ij}=0$ otherwise. A row stochastic matrix has at least one eigenvalue equal to $1$. In particular, $a_{ij}>0$ if $(j,i)\in\Ecal$ and $a_{ij}=0$ otherwise~\cite{HornJohnson85}. 

\section{Problem Formulation}\label{sec:PF}

\begin{figure}
\begin{center}
\includegraphics[width=.4\textwidth]{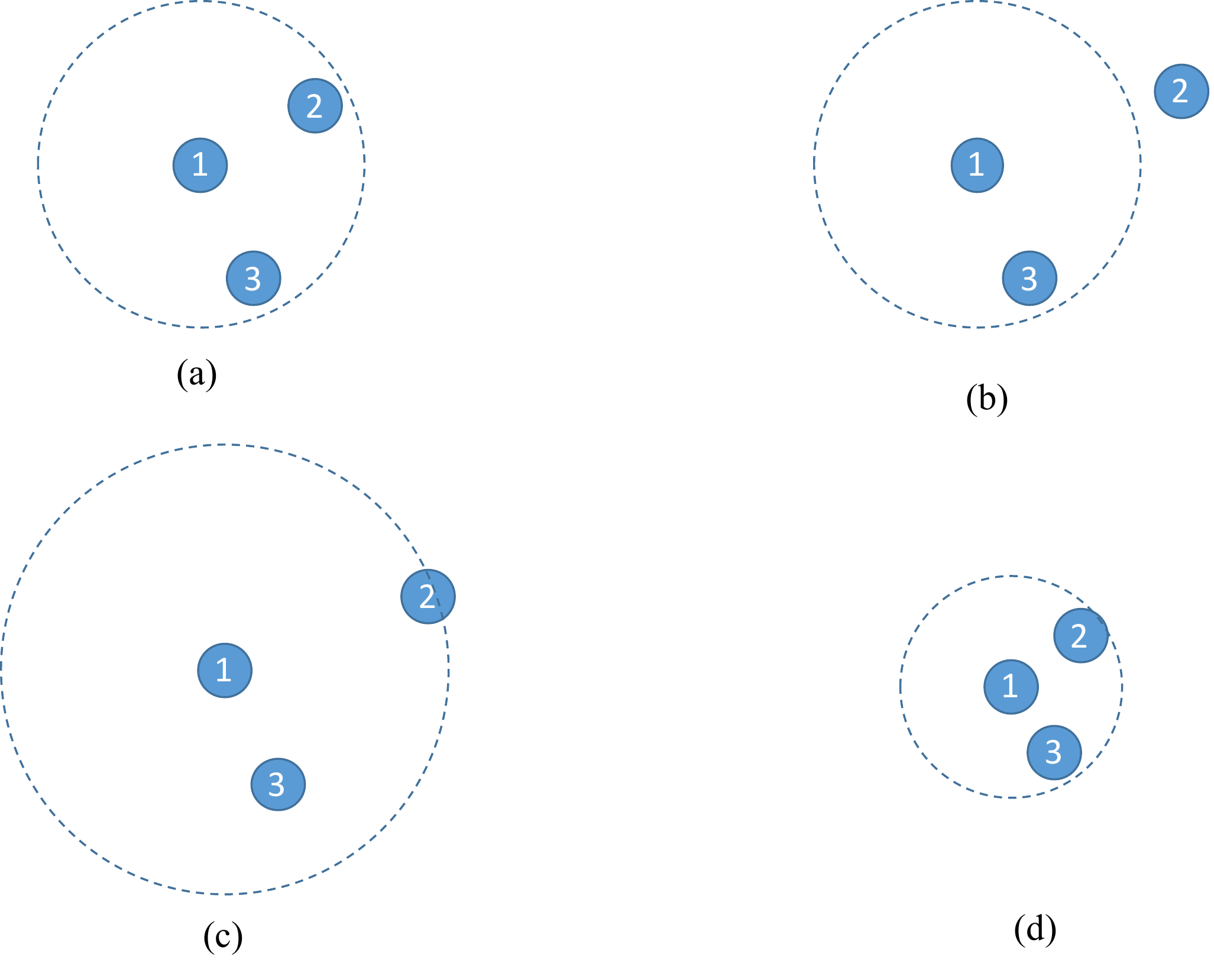}    
\caption{The impact of variable communication disks on network topology. The dashed circle represents the communication range of agent $1$.  The subfigure (a) shows the original communication range when agent $1$ can send its information to both agents $2$ and $3$. The subfigure (b) shows the loss of information transmission from agent $1$ to agent $2$ due to the increased distance from agent $1$ and $2$. The subfigure (c) shows agent $1$ can send its information to both agents $2$ and $3$ given their relative locations via changing the communication range. The subfigure (d) shows agent $1$ can send its information to both agents $2$ and $3$ with its communication range less than the original one in subfigure (a).}  
\label{fig:vd}                                 
\end{center}                                 
\end{figure}

Fig.~\ref{fig:vd} demonstrates how variable communication ranges will affect the network topology for a team of three agents. Given the initial communication range for agent $1$, it can send its information to the other two agents, as shown in Fig.~\ref{fig:vd}(a). However, if agent $2$ moves far away, agent $1$ loses its communication to agent $2$ if the communication range remains the same for agent $1$, as shown in Fig.~\ref{fig:vd}(b). By increasing the communication range, agent $1$ can regain its communication with agent $2$, as shown in Fig.~\ref{fig:vd}(c). Finally, when agents $2$ and $3$ get closer to agent $1$, a smaller communication range can be assigned to agent $1$ that can still maintain required communication pattern in Fig.~\ref{fig:vd}(a), as shown in Fig.~\ref{fig:vd}(d). One interesting question we try to answer in this paper is: can we design proper network topology control technique, based on varying communication ranges, such that desired network topology is maintained with less communication energy consumption? 

In this paper, we consider the distributed network topology control problem for multi-agent systems in consensus missions. The \textit{objective} is to design local topology control algorithms such that networked agents can reach agreement on their final states via designing their variable communication ranges appropriately. The goal is to adjust the communication ranges such that a desired connectivity property is guaranteed for the desired consensus behavior. This paper will address the case when each agent is described by single-integrator kinematics.

We here consider the problem that a team of $N$ networked agents with dynamics given by
\begin{align}\label{eq:1st}
r_i[k+1] =r_i[k]+Tu_i[k], \quad i=1,\cdots,N
\end{align}
where $r_i\in\re^2$ is the location of the $i$th agent in the 2D space, $u_i$ is the control input to be designed for the $i$th agent, $T< \frac{1}{N}$ is the sampling period, and $k$ is the time step index. In the common wireless network model, the power needed to transmit data from one agent $i$ to another agent $j$ is proportional to their Euclidean distance $\norm{r_i-r_j}^\alpha$, where $\alpha$ is a constant that varies within the interval $[2,4]$~\cite{WangHemsteadYang06}. In other words, each agent can send data to its neighbors up to the distance $d$ with transmission power proportional to $d^\alpha$. Let $d_i[k]$ be the communication range for the $i$th agent at time step $k$. The objective is to design $u_i[k]$ and $d_i[k]$ for each $i$, based on $r_i[k]$ and $r_j[k],~j\in \Ncal_i[k]$, such that 
\begin{align}
r_i[k]-r_j[k]=0,\quad k\to\infty.
\end{align}


The existing research only addresses the issue when continuous-time dynamics were considered. The consideration of continuous-time dynamics allows the redesign of consensus control algorithms such that that the control input will push agents closer if they are close to the communication limit. Such a controller design technique requires continuous communication in order to timely monitor the distance between a pair of agents. In a discrete-time setting, such a technique fails to work because the inter-agent distance cannot be monitored continuously. The proposed new control technique, based on distributed network topology control, can solve the two problems by properly designing communication ranges.

\section{Network Topology Control with Variable Communication Ranges for Multi-agent Consensus}\label{sec:consensus_1st_order}

In this section, we consider the case when the agent dynamics are given by~\eqref{eq:1st}. We first analyze how to design $d_i$ such that the desired connectivity condition can be ensured by using variable communication ranges. Then the network topology control technique will be leveraged with the existing consensus control algorithms to solve the well-known consensus problem when each agent has limited but variable communication ranges. Finally, we will analyze the energy consumption and compare it with the traditional approach when each agent has fixed and common communication range.

\begin{figure}
\begin{center}
\includegraphics[width=.2\textwidth]{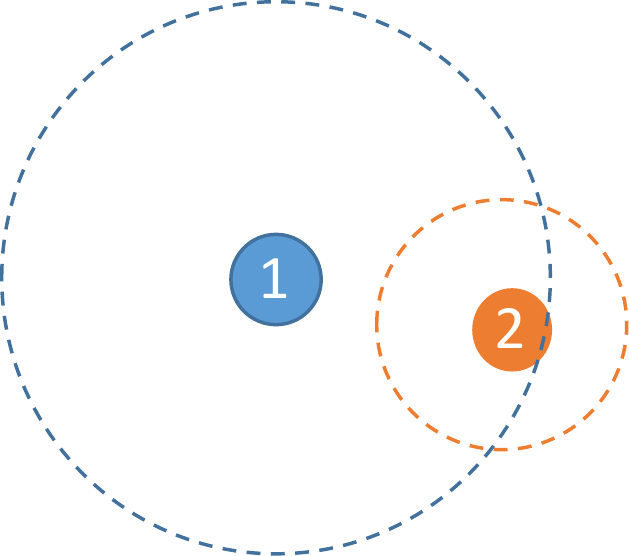}    
\caption{An example of incoming neighbors and outgoing neighbors.}  
\label{fig:in_on}                                 
\end{center}                                 
\end{figure}

Let the communication range for an agent, labeled as $i$, be given by $d_i[k]$ at the time step $k$. Then this agent can send its information to other agents whose Euclidean distances from the agent $i$ is not larger than $d_i[k]$. Mathematically, we describe the instantaneous \textit{outgoing} neighbors for agent $i$ as 
\begin{align}\label{eq:neighbor}
\Ncal^O_i[k]=\{j|\norm{r_i[k]-r_j[k]}\leq d_i[k],~j\in\{1,\cdots,N\}\setminus\{i\}\}.
\end{align}
For agent $i$, we describe its instantaneous \textit{incoming} neighbors as
\begin{align}\label{eq:neighbor}
\Ncal^I_i[k]=\{j|\norm{r_i[k]-r_j[k]}\leq d_j[k],~j\in\{1,\cdots,N\}\setminus\{i\}\}.
\end{align}
The difference between $\Ncal^O_i[k]$ and $\Ncal^I_i[k]$ in their definitions is that $\norm{r_i[k]-r_j[k]}\leq d_i[k]$ means that agent $j$ is within the communication range of agent $i$ while $\norm{r_i[k]-r_j[k]}\leq d_j[k]$ means that agent $i$ is within the communication range of agent $j$. For example, Fig.~\ref{fig:in_on} is an example demonstrating the difference between incoming neighbors and outgoing neighbors. For agent $1$, its incoming neighbor is empty while its outgoing neighbor is agent $2$. For agent $2$, its incoming neighbor is agent $1$ while its outgoing neighbor is empty. For agent $i$, its incoming edges and outgoing edges are then defined as
\begin{align}\label{eq:neighbor}
\Ecal^O_i[k]=\{(i,j)|\norm{r_i[k]-r_j[k]}\leq d_i[k],~j\in\{1,\cdots,N\}\setminus\{i\}\}.
\end{align}
and
\begin{align}\label{eq:neighbor}
\Ecal^I_i[k]=\{(j,i)|\norm{r_i[k]-r_j[k]}\leq d_j[k],~j\in\{1,\cdots,N\}\setminus\{i\}\}.
\end{align}
Define $\Ecal^O[k]\defeq \bigcup_{i=1}^N \Ecal^O_i[k]$ and $\Ecal^I[k]\defeq \bigcup_{i=1}^N \Ecal^I_i[k]$. Then we have the following property regarding $\Ecal^O[k]$ and $\Ecal^I[k]$.

\begin{lemma}\label{lem:equivalence}
$\Ecal^O[k]\equiv\Ecal^I[k]$ for any time step $k$.
\end{lemma}
\proof Note that each edge $(m,n)$ can be uniquely represented in the form of $\norm{r_m[k]-r_n[k]}\leq d_n[k]$ and $\norm{r_m[k]-r_n[k]}\leq d_m[k]$. The set $\Ecal^O[k]$ is given by
$$\Ecal^O[k]=\{(m,n)|\norm{r_m[k]-r_n[k]}\leq d_m[k],\\~m,n\in\{1,\cdots,N\}\}$$
By changing variables, \textit{i.e.}, $m\to n$ and $n\to m$, it follows that 
\begin{align*}
&\{(m,n)|\norm{r_m[k]-r_n[k]}\leq d_m[k],~m,n\in\{1,\cdots,N\}\}\\
=&\{(n,m)|\norm{r_n[k]-r_m[k]}\leq d_n[k],~m,n\in\{1,\cdots,N\}\}\\
=&\{(n,m)|\norm{r_m[k]-r_n[k]}\leq d_n[k],~m,n\in\{1,\cdots,N\}\}\\
=&\Ecal^I[k].
\end{align*}
Therefore, $\Ecal^O[k]$ and $\Ecal^I[k]$ are always equivalent.
\endproof

Lemma~\ref{lem:equivalence} shows that the network topology can be equivalently described by $(\Vcal,\Ecal^O[k])$ and $(\Vcal, \Ecal^I[k])$. To preserve all connectivity patterns, \textit{i.e.,} all edges, two different methods can be adopted. The first approach is to adjust the control input for each agent such that all edges in $\Ecal_i^I[k]$ are preserved. Because each agent cannot control the communication ranges of its incoming neighbors $\Ncal_i^I[k]$, it has to adjust its control input properly. Such a connectivity maintenance method has been developed for continuous-time systems via designing control algorithms based on potential functions~\cite{JiEgerstedt07,ZavlanosPappas07,ZavlanosPappas08,DimarogonasKyriakopoulos08,SuWangChen10,CaoRen12,KanDSD12,PoonawalaSES15}. The second approach is to adjust the communication range for each agent such that all edges in $\Ecal_i^O[k]$ are preserved. Since each agent has no control of how its outgoing neighbors $\Ncal_O^I[k]$ will adjust their control inputs, it has to adjust its communication ranges to guarantee that all edges in $\Ecal_i^O[k]$ be preserved.

To ensure that all edges in $\Ecal_i^O[k]$ can be preserved, agent $i$ needs to predict how its outgoing neighbors will behave. The existing consensus control algorithm given by
\begin{align}\label{eq:consensus-existing}
u_i[k]=-\sum_{j\in\Ncal^I_i[k]} (r_i[k]-r_j[k])
\end{align}
needs to be redesigned because the control input of each neighbor of agent $i$, denoted by $u_j[k], j\in\Ncal_i^O[k]$, is determined by the incoming neighbors of agent $j$. Because $\norm{u_j[k]}$ can be arbitrarily large, the outgoing neighbors of agent $i$ could escape from agent $i$ arbitrarily fast. By revising~\eqref{eq:consensus-existing}  as
\begin{align}\label{eq:consensus-bounded}
u_i[k]=-\text{sat}\left[\sum_{j\in\Ncal^I_i[k]} (r_i[k]-r_j[k])\right]
\end{align}  
where $\text{sat}(\cdot)$ is a saturation function defined as
\begin{align*}
\text{sat}(z) = \left\{
\begin{array} {ll}
z,&\norm{z}<=\gamma,\\
\gamma \frac{z}{\norm{z}},&\text{otherwise},
\end{array}\right.
\end{align*}
where $\gamma$ is a positive constant representing the upper bound of the control input.
The saturation function can guarantee that the control input be always bounded, and thus the action of each agent can be predicted. Note also that the control inputs for physical agents are always bounded. The following lemma shows that how $r_i[k]-r_j[k]$ will evolve given the control algorithm~\eqref{eq:consensus-bounded}.

\begin{lemma}\label{lem:state-difference-prediction}
If one agent $i$ can communicate with agent $j$ at step $k$, then their distance can grow at most $(\norm{u_i[k]}+\gamma)T$.
\end{lemma}
\proof From~\eqref{eq:1st} and~\eqref{eq:consensus-bounded}, we can obtain that
\begin{align*}
r_i[k+1]=r_i[k]-T\text{sat}\left[\sum_{j\in\Ncal^I_i[k]} (r_i[k]-r_j[k])\right]
\end{align*} 
and 
\begin{align*}
r_j[k+1]=r_j[k]-T\text{sat}\left[\sum_{\ell\in\Ncal^I_j[k]} (r_j[k]-r_\ell[k])\right].
\end{align*} 
It then follows that
\begin{align*}
&\norm{r_i[k+1]-r_j[k+1]}\\
\leq &\norm{r_i[k]-r_j[k]}+T\norm{\text{sat}\left[\sum_{\ell\in\Ncal^I_j[k]} (r_j[k]-r_\ell[k])\right]}\\
&+T\norm{\text{sat}\left[\sum_{\ell\in\Ncal^I_j[k]} (r_j[k]-r_\ell[k])\right]}\\
\leq &\norm{r_i[k]-r_j[k]}+(\norm{u_i[k]}+\gamma)T
\end{align*}
where we used the fact that $\norm{u_j[k]}\leq \gamma$ due to the introduction of saturation function in~\eqref{eq:consensus-bounded}.
\endproof

With the aid of Lemma~\eqref{lem:state-difference-prediction}, we have the following lemma regarding the connectivity control for networked multi-agent system with dynamics given by~\eqref{eq:1st}.  

\begin{lemma}\label{lem:maintain}
For a team of multi-agent systems with dynamics given by~\eqref{eq:1st} with the control input designed as~\eqref{eq:consensus-bounded}, if the communication ranges $d_i[k+1]$ is chosen as $\max_{j\in\Ncal^O_i[k]}\norm{r_i[k]-r_j[k]}+(\norm{u_i[k]}+\gamma)T$, the connectivity patterns can be always preserved.
\end{lemma}
\proof We prove the theorem by induction. When $k=0$, it follows that
\begin{align*}
d_i[1]=\max_{j\in\Ncal^O_i[0]}\norm{r_i[0]-r_j[0]}+(\norm{u_i[0]}+\gamma)T.
\end{align*} 
In other words, we have that 
\begin{align*}
d_i[1]\geq \norm{r_i[0]-r_j[0]}+(\norm{u_i[0]}+\gamma)T, \forall j\in\Ncal^O_i[0].
\end{align*} 
According to Lemma~\ref{lem:state-difference-prediction}, for each $j\in\Ncal^O_i[0]$, we can obtain that $j\in\Ncal^O_i[1]$. Let $j\in\Ncal^O_i[k]$ hold for some $k$. By following a similar analysis, it can be obtained that $j\in\Ncal^O_i[k+1]$. Therefore, the lemma holds true.
\endproof

According to Lemma~\ref{lem:maintain}, network topology can be effectively controlled locally by changing communication ranges. By selecting the communication ranges as described in Lemma~\ref{lem:maintain}, we have the following theorem regarding consensus for agents with single-integrator kinematics.

\begin{theorem}\label{th:consensus1}
For a team of $N$ agents with dynamics given by~\eqref{eq:1st}, the control algorithm~\eqref{eq:consensus-bounded} can guarantee consensus, \textit{i.e.,} $r_i[k]-r_j[k]\to 0$ as $k\to\infty$, when each agent has a limited by variable communication range given by $\max_{j\in\Ncal^O_i[k]}\norm{r_i[k]-r_j[k]}+(\norm{u_i[k]}+\gamma)T$ if the initial interaction graph $\Gcal[0]=(\Vcal, \Ecal^O[0])$ has a directed spanning tree. 
\end{theorem}
\proof By letting $\max_{j\in\Ncal^O_i[k]}\norm{r_i[k]-r_j[k]}+(\norm{u_i[k]}+\gamma)T$, it follows from Lemma~\ref{lem:maintain} that the connectivity patterns can be always preserved. When the initial interaction graph $\Gcal[0]$ has a directed spanning tree, it then follows that the interaction graph $\Gcal[k],~k=1,\cdots,$ has a directed spanning tree. It then follows from~\cite{RenBeard05_TAC} that $r_i[k]-r_j[k]\to 0$ as $k\to\infty$.  
\endproof

Theorem~\ref{th:consensus1} shows that consensus can be achieved for networked multi-agent systems in the discrete-time setting. In particular, we consider the general case when the network topology is directed and switching when each agent has limited but varying communication ranges. The main idea is to change the communication range such that the existing communication patterns can be always preserved. When consensus is reached, it can be observed that the communication range $\max_{j\in\Ncal^O_i[k]}\norm{r_i[k]-r_j[k]}+(\norm{u_i[k]}+\gamma)T$ becomes $\gamma T$ because the state difference $r_i[k]-r_j[k]=0$ and the control input $u_i[k]=0$. Clearly, requiring a constant communication range $\gamma T$ is a waste of communication power. To further reduce the communication power consumption, we proposed a modified distributed communication range control strategy, as described in Algorithm~\ref{alg:range-new}. 

\begin{algorithm}
\caption{Modified communication range control}\label{alg:range-new}
\begin{algorithmic}[1]
\State{$d_i[k+1]\gets 0$}
\State Compute $u_i[k]$ according to~\eqref{eq:consensus-bounded}
\If {$\Ncal_i^O[k]\cup \{i\}\neq \Vcal$} 
\State $d_i[k+1] \gets \max_{j\in\Ncal^O_i[k]}\norm{r_i[k]-r_j[k]}\newline~~~~~~~~~~~~~~~~~~~~~+\norm{u_i[k]}+T\gamma$;
\EndIf
\If {$\Ncal_i^O[k]\cup \{i\}\equiv \Vcal$}
\State $d_i[k+1] \gets 2\max_j \norm{r_i[k]-r_j[k]}$;
\EndIf
\State return $d_i[k+1]$
\end{algorithmic}
\end{algorithm}

We have the following lemma regarding the new communication range control algorithm~\ref{alg:range-new}.

\begin{lemma}\label{lem:maintain-new}
For a team of multi-agent systems with dynamics given by~\eqref{eq:1st} with the control input designed as~\eqref{eq:consensus-bounded}, if the communication ranges $d_i[k+1]$ is chosen as described in Algorithm~\ref{alg:range-new}, the connectivity patterns can be always preserved.
\end{lemma}
\proof We prove the lemma by considering two cases: (1) $\Ncal_i^O[k]\cup \{i\}\neq \Vcal$; and (2) $\Ncal_i^O[k]\cup \{i\}\equiv \Vcal$. We will show that the connectivity patterns can be preserved for both cases.

Case (1): $\Ncal_i^O[k]\cup \{i\}\neq \Vcal$. In this case, the set of the outgoing agents of agent $i$ and the agent $i$ itself does not contain all possible agents. In other words, there exists at least one agent that is not an outgoing neighbor of agent $i$. According to Algorithm~\ref{alg:range-new}, the communication range is updated by the strategy described in Lemma~\ref{lem:maintain}. It then follows from Lemma~\ref{lem:maintain} that communication patterns can be preserved. 

Case (2): $\Ncal_i^O[k]\cup \{i\}\equiv \Vcal$. In this case, the set of the outgoing agents of agent $i$ and the agent $i$ itself contains all possible agents. Therefore, the agent $i$ can send its information to all other agents at the time step $k$. Define 
\begin{align*}
&C(r_i[k],\max_j \norm{r_i[k]-r_j[k]})\\
\defeq &\{x|\norm{x-r_i[k]}\leq \max_j \norm{r_i[k]-r_j[k]}\}.
\end{align*} 
It then follows that $r_j[k]\in C(r_i[k],\max_j \norm{r_i[k]-r_j[k]}),~\forall j=1,\cdots,N$. By using the control algorithm~\eqref{eq:consensus-bounded}, each agent will move towards its incoming neighbors. Therefore, all agents at time step $k+1$ will be inside the convex set formed by all agents at time step $k$~\cite{Moreau05}. Because the convex set formed by all agents is contained in the set $C(r_i[k],\max_j \norm{r_i[k]-r_j[k]})$, all agents at the time step $k+1$ will remain in the set $C(r_i[k],\max_j \norm{r_i[k]-r_j[k]})$. Therefore, we have that $r_j[k+1]\in C(r_i[k],\max_j \norm{r_i[k]-r_j[k]}),~\forall j=1,\cdots,N$. When all agents are in the set $C(r_i[k],\max_j \norm{r_i[k]-r_j[k]})$, the maximum distance among any pair of agents is no larger than $2\max_j \norm{r_i[k]-r_j[k]}$ at the next time step $k+1$. When $d_i[k+1] \gets 2\max_j \norm{r_i[k]-r_j[k]}$ as described in Algorithm~\ref{alg:range-new}, agent $i$ can send its information to all other agents at the time step $k+1$.

Because Cases (1) and (2) contain all possible communication graphs associated with the $N$ agents at the time step $k$, it then follows from the previous analysis in the proof that the connectivity patterns can be always preserved.
\endproof

Since the existing communication patterns can be preserved when designing communication ranges based on Algorithm~\ref{alg:range-new}, we have the following results regarding consensus for agents with single-integrator kinematics.
\begin{corollary}\label{co:consensus2}
For a team of $N$ agents with dynamics given by~\eqref{eq:1st}, the control algorithm~\eqref{eq:consensus-bounded} can guarantee consensus, \textit{i.e.,} $r_i[k]-r_j[k]\to 0$ as $k\to\infty$, when each agent has a limited by variable communication range selected based on Algorithm~\ref{alg:range-new} if the initial interaction graph $\Gcal[0]=(\Vcal, \Ecal^O[0])$ has a directed spanning tree. 
\end{corollary}
\proof The proof is similar to the one of Theorem~\ref{th:consensus1}.
\endproof

Compared with the communication range control strategy in Theorem~\ref{th:consensus1}, the strategy proposed in Algorithm~\ref{alg:range-new} can potentially save a significant amount of communication energy, especially when all agents are close to each other. In particular, the requested communication range $d_i[k]\geq T\gamma$ for the communication range control strategy in Theorem~\ref{th:consensus1} even if consensus is reached. However, the requested communication range $d_i[k]\to 0$ if consensus is reached. 

In the previous part of this section, we assume that each agent will send its information to its outgoing neighbors at each time step. This assumption can be further relaxed by letting each agent send its information to its outgoing neighbors intermittently. The following lemma presents a general extension to the Lemma~\ref{lem:maintain}.

\begin{lemma}\label{lem:maintain2}
Consider a team of multi-agent systems with dynamics given by~\eqref{eq:1st} with the control input designed as~\eqref{eq:consensus-bounded}. Let agent $i$ send its information to its outgoing neighbors at $\kappa^i_1, \kappa^i_2, \cdots$. If the communication ranges $d_i[\kappa^i_{s+1}]$ is chosen as $\max_{j\in\Ncal^O_i}\norm{r_i[{\kappa^i_s}]-r_j[{\kappa^i_s}]}+(\kappa^i_{s+1}-\kappa^i_s)(\norm{u_i[{\kappa^i_s}]}+\gamma)T$, the connectivity patterns at the time step $s$ can be always preserved.
\end{lemma}
\proof By considering $\kappa^i_{s+1}-\kappa^i_s$ as the new sampling period $T$, it then follows directly from the proof of Lemma~\ref{lem:maintain} that the conclusion in this lemma holds.
\endproof

Note that the proofs of both Lemma~\ref{lem:maintain} and Lemma~\ref{lem:maintain2} do not rely on the synchronous communication because each agent only needs to maintain the communication from itself to its outgoing neighbors. Therefore, asynchronous communication can be used to preserve connectivity patterns, when each agent can independently plan when it will send its information to its outgoing neighbors. 

Although we discussed the possibility to preserve connectivity by using variable communication ranges, it is unclear whether more communication energy is needed. The power consumption from one sensor to another sensor is typically determined by their distance~\cite{WangHemsteadYang06}. In particular, by excluding the power consumption at the circuit level, a general model for the power consumption can be mathematically described as~\cite{WangHemsteadYang06}
\begin{align}\label{eq:comm-power}
P(d) = \epsilon d^{\alpha},
\end{align}
where $P(d)$ is the power consumption, $d$ is the communication range, $\epsilon$ is a positive constant, and $\alpha \in[2,4]$ is also a positive constant. Based on this practical communication energy consumption model, we now present the following theorem that illustrates the relationship between the power consumption using fixed and common communication ranges and that using variable communication ranges as described in Algorithm~\ref{alg:range-new}.   

\begin{theorem}\label{th:power}
For a team of $N$ agents with dynamics given by~\eqref{eq:1st}, assume that the initial interaction graph $\Gcal[0]=(\Vcal, \Ecal^O[0])$ has a directed spanning tree. Then the control algorithm~\eqref{eq:consensus-bounded} with variable communication range control given in Algorithm~\ref{alg:range-new} can guarantee consensus with much less communication energy consumption than the case when a fixed and common communication range is used for all agents. In addition, the total communication energy consumption using Algorithm~\ref{alg:range-new} is always finite.
\end{theorem}
\proof When the initial interaction graph $\Gcal[0]=(\Vcal, \Ecal^O[0])$ has a directed spanning tree, it then follows from Corollary~\ref{co:consensus2} that consensus is reached. In other words, $\norm{r_i[k]-r_j[k]}\to 0$ as $k\to\infty$. Therefore, for each agent $i$, there must exist a positive integer, $t^i_1$, such that 
$$\max_{i,j\in\{1,\cdots, N\}} \norm{r_i[k]-r_j[k]}\leq T\gamma,~\forall k\geq t^i_1$$
According to Algorithm~\ref{alg:range-new}, the communication range is given by $2\max_j \norm{r_i[k]-r_j[k]}$ for all $k\geq t^i_1$. Because consensus is reached as $k\to\infty$, for each agent $i$, there exists another positive integer, $t^i_2(\sigma)$, such that $2\max_j \norm{r_i[k]-r_j[k]}\leq \sigma$, where $\sigma$ is an arbitrarily small positive number. Let $P_i[k]$ be the communication power consumption at the time step $k$. Then $P_i[k]\leq \epsilon \sigma^\alpha$ for all $k\geq t^i_2(\sigma)$. Therefore, the overall communication power consumption $P_{p}$ under the proposed communication range control strategy, described in Algorithm~\ref{alg:range-new}, satisfies 
\begin{align}\label{eq:proposed-power}
P_{p} \leq \sum_{i=1}^N\left(\sum_{k=1}^{t^i_2-1} P_i[k] +\sum_{k=t^i_2}^{\infty} \epsilon \sigma^\alpha\right).
\end{align}  

For the existing communication strategy used in solving consensus problems that assumes a fixed and common communication range, denoted by $\delta$, the overall communication power consumption $P_{f}$ is given by 
\begin{align}\label{eq:old-power}
P_{f} = N\sum_{k=t_1}^{\infty} \epsilon \delta^\alpha.
\end{align}  
Since $\sigma$ can be chosen arbitrarily small, $\delta>\sigma$ if $t^i_2(\sigma)$ is chosen properly. By comparing~\eqref{eq:proposed-power} and~\eqref{eq:old-power}, it can be obtained that $P_f>P_p$. Therefore, the proposed Algorithm~\ref{alg:range-new} requires less communication power.

We now show that the total communication energy consumption using Algorithm~\ref{alg:range-new} is finite. Because consensus is guaranteed using the proposed control algorithm~\eqref{eq:consensus-bounded}, there exists a time step $t$ such that 
$$u_i[k]=-\sum_{j\in\Ncal^I_i[k]} (r_i[k]-r_j[k]), \quad k\geq t.$$
In other words, the control input $u_i[k]$ satisfies the property $\norm{u_i[k]}\leq \gamma$. Therefore, the closed-loop system of~\eqref{eq:1st} using~\eqref{eq:consensus-bounded} becomes a linear system given by
\begin{equation}\label{eq:closed}
r_i[k+1]=r_i[k]-T\sum_{j\in\Ncal^I_i[k]} (r_i[k]-r_j[k]), \quad k\geq t.
\end{equation}
Let $\bar{t}\defeq \max\{t,t^i_1,i=1,\cdots,N\}$. Then~\eqref{eq:closed} becomes
\begin{equation}\label{eq:closed}
r_i[k+1]=r_i[k]-T\sum_{j=1}^N (r_i[k]-r_j[k]), k\geq \bar{t}
\end{equation}
because each agent can send its information to all others when $k\geq \max\{t^i_1,i=1,\cdots,N\}$.
When $T<\frac{1}{N}$, it follows that~\eqref{eq:closed} can be rewritten in a matrix form as
\begin{equation}\label{eq:closed-matrix}
r[k+1]=(I_N-T\Lcal)r[k], k\geq \bar{t}
\end{equation}
where $r=[r_1,\cdots,r_N]^T$ and $\Lcal$ is the Laplacian matrix associated with a complete graph for the $N$ agents. By selecting $T<N$, $(I_N-T\Lcal)$ is a stochastic matrix with positive diagonal entries. Therefore, there exists a positive constant $\beta\in(0,1)$ such that
\begin{align*}
&\max_{i,j\in\{1,\cdots,N\}} \norm{r_i[k+1]-r_j[k+1]}\\
\leq&\beta\max_{i,j\in\{1,\cdots,N\}} \norm{r_i[k]-r_j[k]}
\end{align*} 
for all $k\geq \max\{t,t^i_1,i=1,\cdots,N\}$. Let $d_{\bar{t}}\defeq \max_{i,j\in\{1,\cdots,N\}}\norm{r_i[\bar{t}]-r_j[\bar{t}]}$. Then the total communication energy consumption can be written as
\begin{align*}
P_{p} \leq &\sum_{i=1}^N\left(\sum_{k=1}^{\bar{t}-1} P_i[k] +\sum_{k=\bar{t}}^{\infty} \epsilon \sigma^\alpha\right)\\
\leq &\sum_{i=1}^N\left(\sum_{k=1}^{\bar{t}-1} P_i[k] +\sum_{k=\bar{t}}^{\infty} \epsilon (\beta^{k-1} d_{\bar{t}})^\alpha\right)\\
=&\sum_{i=1}^N\sum_{k=1}^{\bar{t}-1} P_i[k]+N\epsilon (d_{\bar{t}})^\alpha \frac{\beta^{\bar{t}-1}}{1-\beta}.
\end{align*}
Therefore, the total communication energy consumption using Algorithm~\ref{alg:range-new} is always finite.
\endproof

In the previous part of this section, we show that a variable communication range control technique yields numerous benefits, including bounded control input, discrete-time communication, and finite communication energy consumption. Note that these benefits can hardly be obtained using the existing potential-based consensus control algorithms. 






\section{Simulation Examples}\label{sec:simulation}
\begin{figure}
\begin{center}
\includegraphics[width=.2\textwidth]{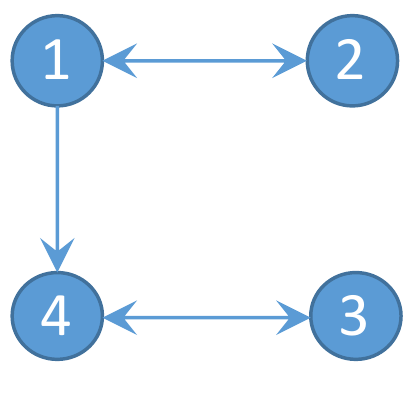}    
\caption{The initial interaction graph $\Gcal[0]$. An arrow from $i$ to $j$ means that agent $i$ can send its information to agent $j$.}  
\label{fig:topology}                                 
\end{center}                                 
\end{figure}

In this section, we will conduct simulation examples to validate the proposed network topology control by using variable communication ranges. We consider a team of $4$ agents in the 2D space. In particular, the sampling period $T=0.1$. The initial states of the fours agents are randomly selected as $r_1[0] = [2,2]$, $r_2[0]= [1.4,3.2]$, $r_3[0]=[3.7,5.2]$, and $r_4[0]=[4.5,4.3]$. The initial communication ranges are selected as $d_1[0] = 3.5$, $d_2[0]=2.5$, $d_3[0]=1.5$, and $d_4[0]=1.4$. The initial communication topology $\Gcal[0]$ is given in Fig.~\ref{fig:topology}. It can be observed from Fig.~\ref{fig:topology} that $\Gcal[0]$ has a directed spanning tree.

\begin{figure}
\begin{center}
\includegraphics[width=.45\textwidth]{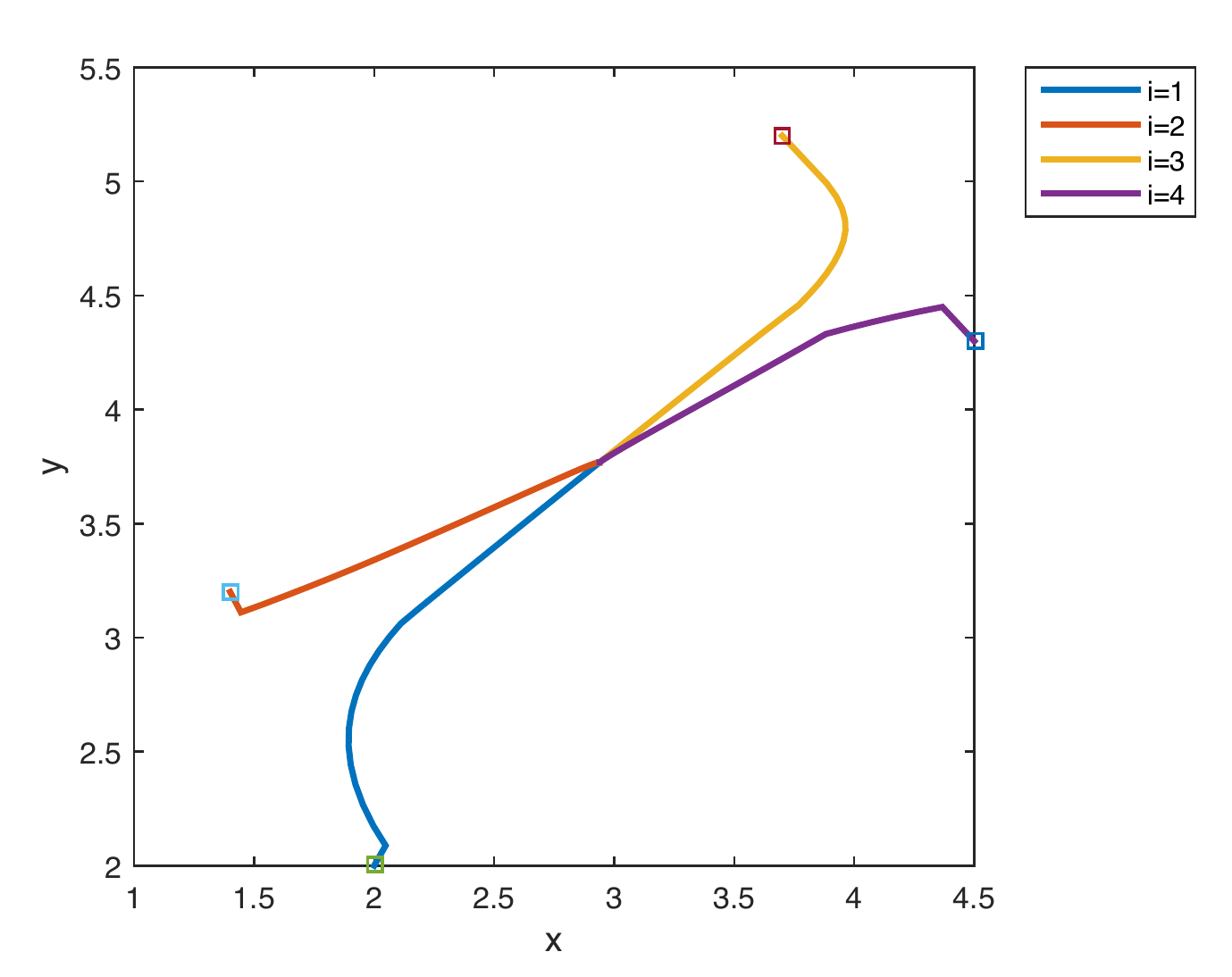}    
\caption{The trajectories of the four agents.}  
\label{fig:traj}                                 
\end{center}                                 
\end{figure}

\begin{figure}
\begin{center}
\includegraphics[width=.45\textwidth]{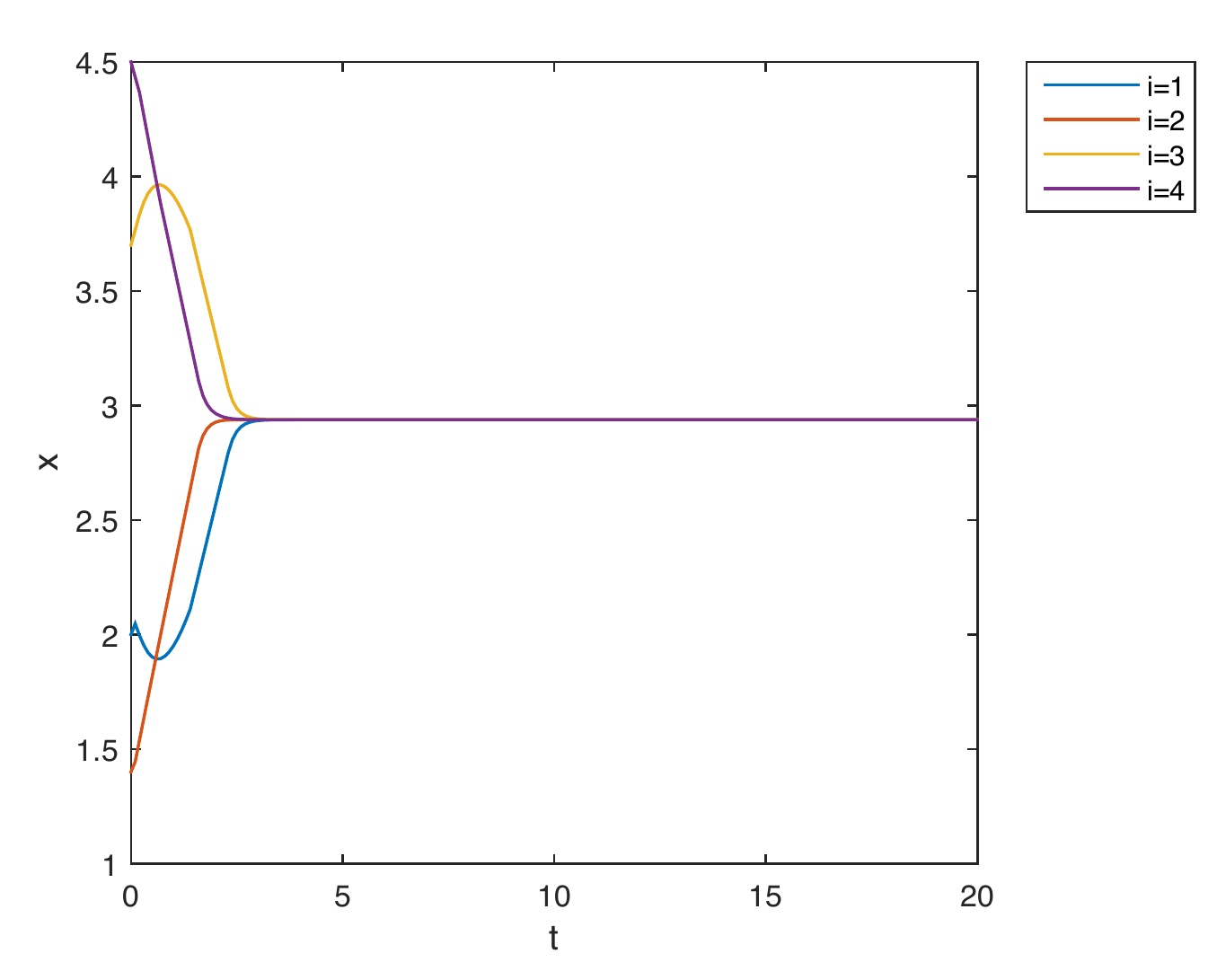}    
\caption{The $x$ components of the trajectories of the four agents.}  
\label{fig:trajx}                                 
\end{center}                                 
\end{figure}

\begin{figure}
\begin{center}
\includegraphics[width=.45\textwidth]{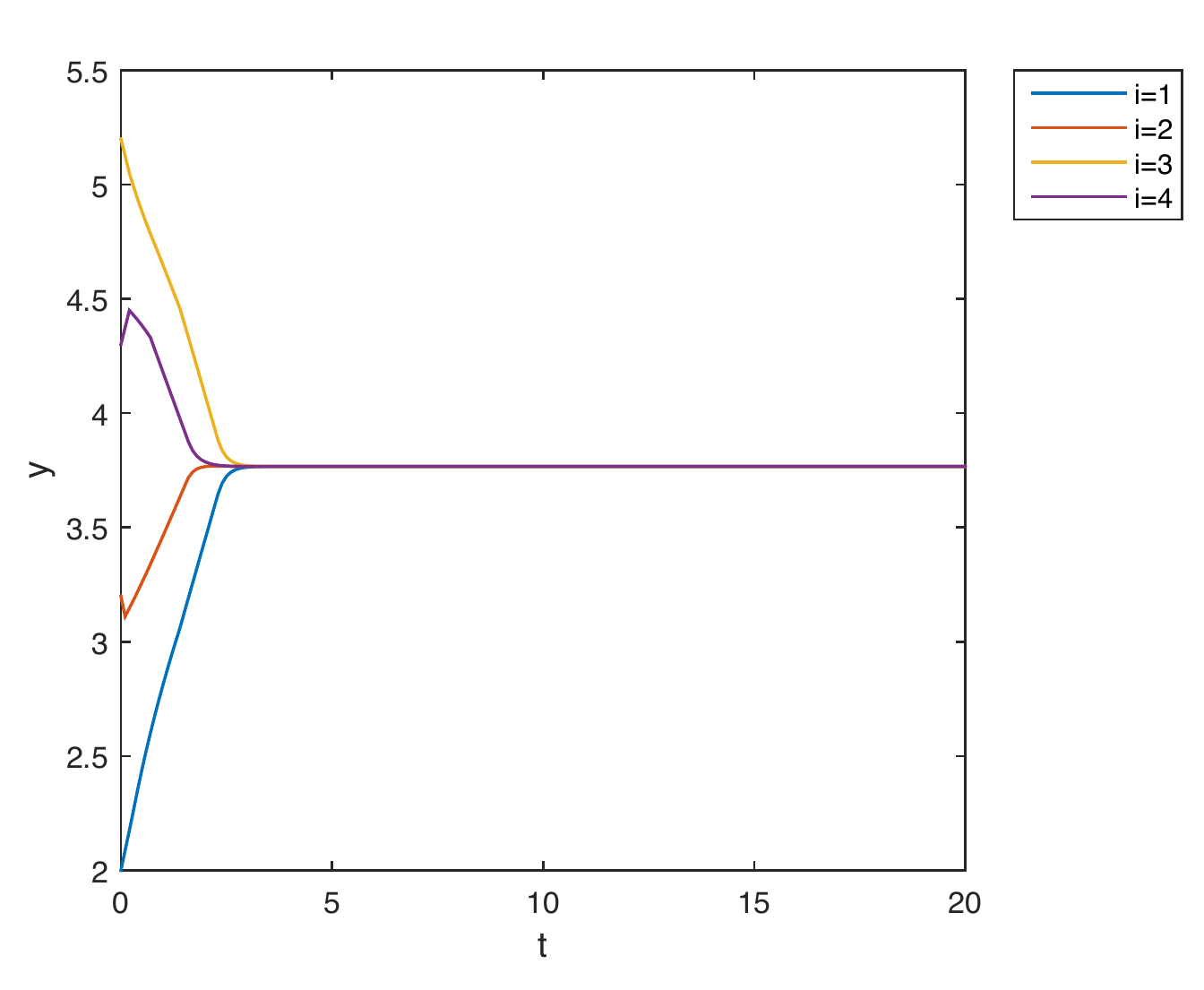}    
\caption{The $y$ components of the trajectories of the four agents.}  
\label{fig:trajy}                                 
\end{center}                                 
\end{figure}

\begin{figure}
\begin{center}
\includegraphics[width=.45\textwidth]{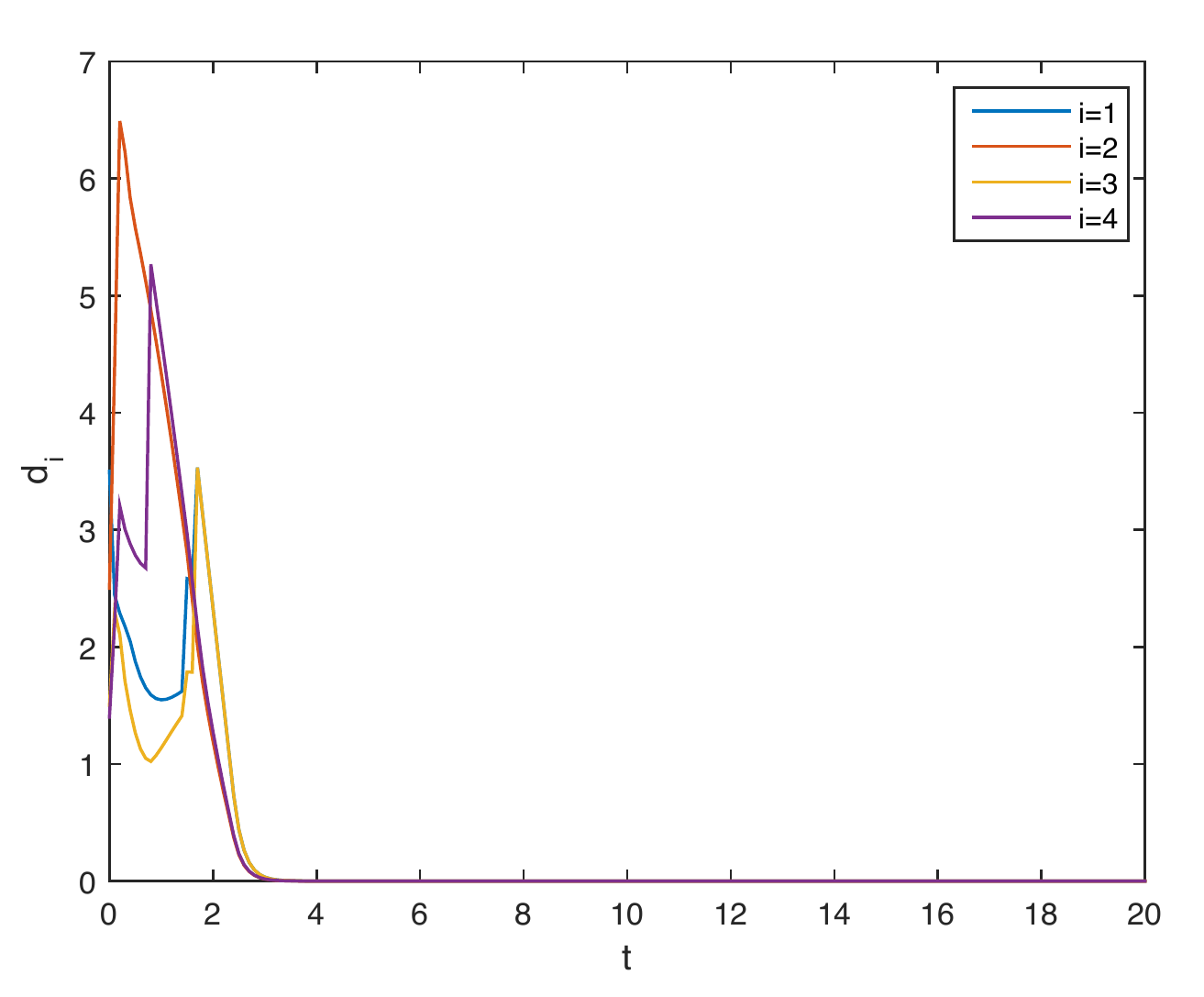}    
\caption{The communication ranges of the four agents.}  
\label{fig:comm_d}                                 
\end{center}                                 
\end{figure}

By using the variable communication range strategy given in Algorithm~\ref{alg:range-new}, Fig.~\ref{fig:traj} shows the trajectories of the four agents using the control algorithm given in~\eqref{eq:consensus-bounded}. Figs.~\ref{fig:trajx} and~\ref{fig:trajy} show, respectively, the $x$ component and the $y$ component of the trajectories of the four agents. It can be seen that the four agents will reach consensus. Fig.~\ref{fig:comm_d} shows how the communicate ranges for the four agents will evolve by using Algorithm~\ref{alg:range-new}. We can observe that the communication ranges will approach zero as the relative state differences among the fours agents converge to zero. Note also that the communication ranges also jump due to the addition of new outgoing neighbors as the four agents move closely to each other.  

\begin{figure}
\begin{center}
\includegraphics[width=.45\textwidth]{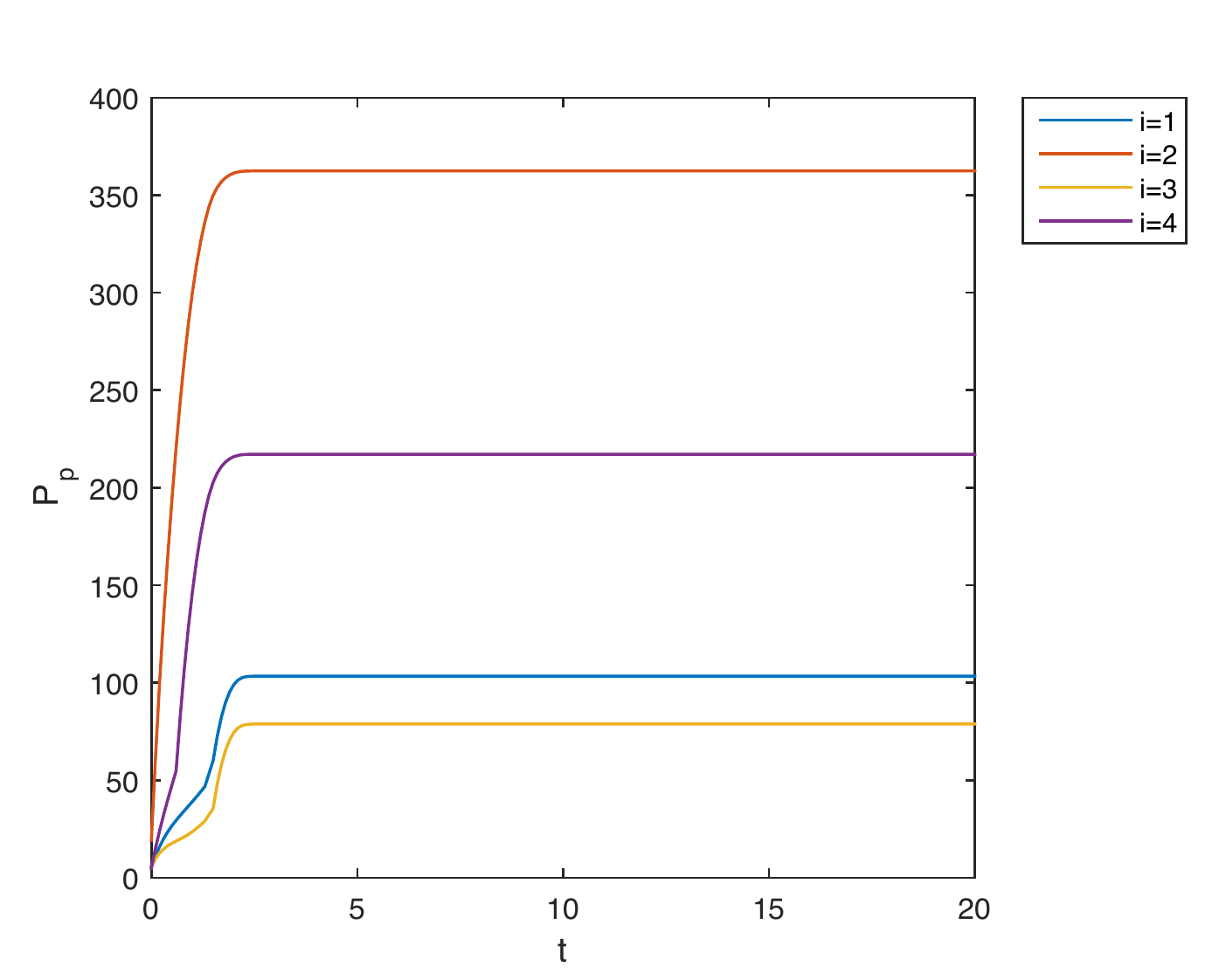}    
\caption{The total communication energy used by the four agents by using Algorithm~\ref{alg:range-new}.}  
\label{fig:energy}                                 
\end{center}                                 
\end{figure}

By letting $\epsilon=1$ and $\alpha=2$, Fig.~\ref{fig:energy} shows the communication energy for each agent using the proposed Algorithm~\ref{alg:range-new}. Fig.~\ref{fig:energy} shows the communication energy for each agent when the initial communication ranges for the four agents remain constant afterwards. It can be seen that the total communication energy using the proposed Algorithm~\ref{alg:range-new} is much smaller than the case when constant communication ranges are used. In particular, the proposed Algorithm~\ref{alg:range-new} requires finite communication energy while the traditional approach requires infinite communication energy. Therefore, the proposed variable communication range technique is more energy-efficient.

\begin{figure}
\begin{center}
\includegraphics[width=.45\textwidth]{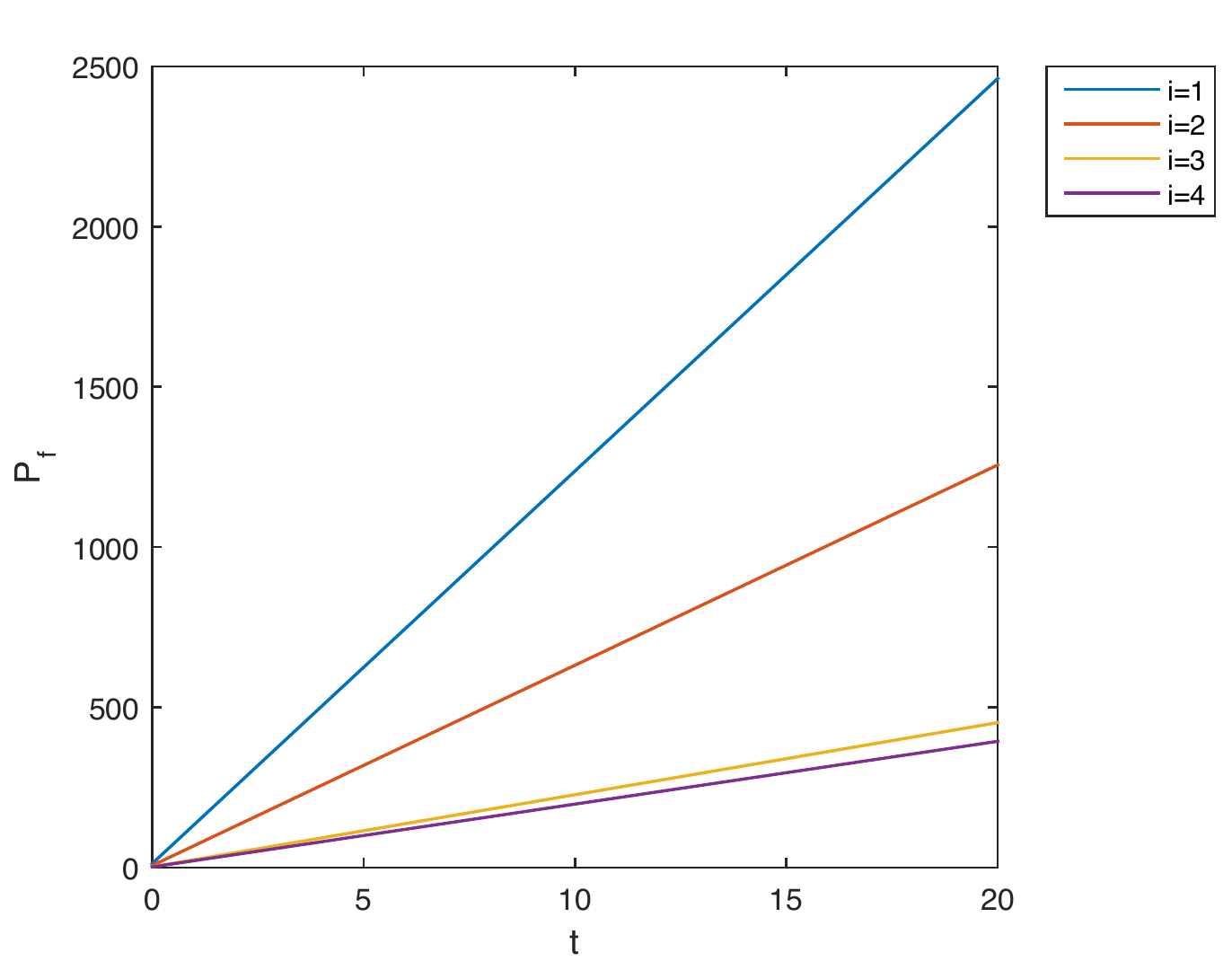}    
\caption{The total communication energy used by the four agents if the initial communication ranges are used for all subsequent times.}  
\label{fig:energy_old}                                 
\end{center}                                 
\end{figure}

\section{Conclusion}\label{sec:conclusion}

In this paper, we considered the network topology control problem for networked cyber-physical systems when each system has limited communication range. Instead of assuming fixed and homogeneous communication ranges, we proposed a new network topology control technique based on variable communication ranges. In particular, for the multi-agent consensus problem, we developed new distributed control algorithms along with variable communication control strategies such that consensus can be reached in the discrete-time setting. In addition, the proposed control algorithms require bounded control inputs with bounded communication energy consumption.

\bibliographystyle{IEEEtran}
\bibliography{refs}
\end{document}